\documentclass[english,pra,twocolumn,groupedaddress,reprint]{revtex4}
\usepackage{graphicx}
\usepackage{amsfonts}
\usepackage[T1]{fontenc}
\usepackage[latin9]{inputenc}
\usepackage{color}
\usepackage{amsmath}
\usepackage{amssymb}
\usepackage{esint}
\usepackage{comment}
\usepackage{bbold}
\usepackage{tikz}
\usepackage{minitoc}
\usepackage{bm}
\usepackage{CJK}

\begin{document}

\title{Classicality and amplification in postselected weak measurement}
\author{Tao Wang\footnote{suiyueqiaoqiao@163.com} $^{1,2}$, Rui Zhang  $^{1}$, Gang Li  $^{3}$, and Xue-Mei Su $^{1}$}
\affiliation{$^{1}$College of Physics, Jilin University, Changchun 130012, People's Republic of China}
\affiliation{$^{2}$College of Physics, Tonghua Normal University, Tonghua 134000, People's Republic of China}
\affiliation{$^{3}$School of Physics and Optoelectronic Technology,
Dalian University of Technology, Dalian 116024, People's Republic of China}
 \date{\today }

\begin{abstract}
Nearly thirty years ago the possibility of anomalous weak amplfication (AWA) was revealed by Aharonov, Albert and Vaidman \cite{Aharonov88}. Recently two papers presents two AWA schemes which are beyond the traditional proposal given by them \cite{Wang15,Howell15}. At the first glance the two papers seems very different. Ref. \cite{Wang15} discusses the thermal light cross-Kerr effect and finds only postselection can give the amplification effect without interference of the postselected meter states, and Ref. \cite{Howell15} shows that only weak interaction itself can give the amplification effect without postselection. Here the relationships between the two papers are pointed out and a generalized framework for AWA via postselecting a pair of orthogonal final states is shown. ~~~~\newline
~~~~\newline
\end{abstract}

\maketitle

In 1988 anomalous weak amplification (AWA) was proposed in the seminal paper given by Aharonov, Albert, and Vaidman \cite{Aharonov88}. A prominent amplification result can be obtained, although weakly disturbing a quantum system given the initial and final states can be only applied. The inconceivable effect was exploited to observe many tiny physical effects \cite{Hosten08,Dixon09}, to directly measure the wavefunction of a quantum system \cite{Bamber11}, and to explain some counter-intuitive quantum phenomena \cite{Aharonov05}. An endeavor to apply AWA into optomechanics was pursued in \cite{Li14,Li15,Li1}, which may be used to detect the faint gravitational wave effect. Reviews on this wonderful idea can be found in \cite{Shikano10,Dresse14}.

This AWA scheme traditionally consists of three steps. \emph{i)} The quantum system to be measured and the meter system to measure are prepared respectively. For simplicity, we choose $|\psi_{i,1}\rangle=\frac{1}{\sqrt{2}}(e^{i\theta}|0\rangle+|1\rangle)$ for the quantum system, and $|\varphi_{i}\rangle=\int dq \varphi(q)|q\rangle$ for the meter system, which is usually a pure Gaussian state. $\theta$ is a small phase shift. \emph{ii)} The quantum system weakly interacts with the meter system via $U=e^{-ig\hat{A}\hat{p}}$. After interaction, the two systems are weakly entangled together, so the state of the whole system is $\int d^{2}q\frac{1}{\sqrt{2}}(e^{i\theta}\varphi(q+g)|0\rangle+\varphi(q-g)|1\rangle)|q\rangle$. Here $\hat{q}$ and $\hat{p}$ are two conjugate variables (called position and momentum), and $\hat{A}|0\rangle=-|0\rangle$, $\hat{A}|1\rangle=+|1\rangle$. \emph{iii)} A final state is particularly selected, which is nearly orthogonal to the initial state. Here we choose $|\psi_{f,1}\rangle=\frac{1}{\sqrt{2}}(|0\rangle-|1\rangle)$ (this postselection is called dark port for small detection probability), so $\langle \psi_{f,1}|\psi_{i,1}\rangle=\frac{1}{2}(e^{i\theta}-1)\approx \frac{\theta}{2}i$ , which is an imaginary value. The final state of the meter system is
\begin{eqnarray}
\int dq\frac{1}{2P}(e^{i\theta} \varphi(q+g)-\varphi(q-g))|q\rangle=\int dq \varphi'(q)dq
\end{eqnarray}
where $P$ is the success detection probability and is very small. Here destructive interference occurs, and we have an amplification effect for the change of the pointer' s mean momentum $p$.

We can also choose $|\psi_{i,2}\rangle=\frac{1}{\sqrt{2}}(t|0\rangle+r|1\rangle)$ for the quantum system as the initial state. $t$ and $r$ are two real numbers close to unitary, and $t^{2}+r^{2}=2$. If postselection is also $|\psi_{f,1}\rangle$, $\langle \psi_{f,1}|\psi_{i,2}\rangle=\frac{1}{2}(t-r)= \frac{\delta}{2}$ is a real value. In this choice we can have an amplification effect for the change of the probe' s mean position $q$, which is mainly concerned with in \cite{Aharonov88} and other early literatures.

Aharonov \emph{et al} explained this AWA via weak value, which is defined as following given the initial state $|\psi_{i}\rangle$ and the final state $|\psi_{f}\rangle$,
\begin{eqnarray}
\langle \hat{A} \rangle_{weak}=\frac{\langle \psi_{f} |\hat{A}|\psi_{i}\rangle}{\langle \psi_{f} |\psi_{i}\rangle}.
\end{eqnarray}
When $|\psi_{i,1}\rangle$ and $|\psi_{f,1}\rangle$ are chosen, $\langle \hat{A} \rangle_{weak,11}\approx -1+\frac{2}{\theta}i$, which has a large imaginary part (imaginal weak value). When $|\psi_{i,2}\rangle$ and $|\psi_{f,1}\rangle$ are selected, $\langle \hat{A} \rangle_{weak,21}= -\frac{t+r}{\delta}$, which is much beyond the eigenvalue spectra of $\hat{A}$ (real weak value), and the absolute value is also very large. Large real weak value corresponds to the large change of the pointer's mean position $q$, and large imaginary weak value can be used to explain the large change of the meter's mean momentum $p$. This relationship has been clarified by Jozsa \cite{Jozsa}. That paper also pointed out that the shift of $p$ is an artifact of postselection.

In our paper the traditional AWA scheme is called AAV scheme (also usually called weak-value amplification). In this scheme, due to the destructive interference in Equ. (1), the state of the meter system can be changed largely. Two years ago Str\"{u}bi and Bruder presented a new scheme (named SB scheme) \cite{Bruder13} via joint weak measurements away from the traditional weak amplification regime. In their paper they pointed out:  ``The crucial information in \emph{imaginary }weak-value amplification is not the average shift of the meter but the correlations induced by the interaction between the system and the meter. These are generally probed by joint measurements on the meter and system.'' Through the maximum-likelihood procedure \cite{Helstrom}, they showed unexpected results that many limits on measurement precision can be removed in this scheme for the special situation of almost output intensities of the two ports. This means two final states $|\psi_{f,2}\rangle=\frac{1}{\sqrt{2}}(|0\rangle+i|1\rangle)$ and $|\psi_{f,3}\rangle=\frac{1}{\sqrt{2}}(|0\rangle-i|1\rangle)$ are both postselected to estimate the small interaction parameter which is different to that in AVA scheme. However the details for this AWA are not clearly clarified.

Generally speaking, postselection, weak entanglement and destructive interference of the postselected meter states (IPMS) are all regarded as the indispensable components for AWA. However Ref. \cite{Wang15} reveals another side of AWA via thermal light cross-Kerr effect: entanglement and IPMS are not necessary for (imaginary) AWA, and postselection plays a classical role. In \cite{Wang15} the initial states of the quantum system and the meter sytem are single-photon entangled state and the thermal state. This is a special and prominent case for the new perspective, and the key thing is that the thermal light is a classical mixed state. In this paper we further explore this meaning to discuss the pure state $|\varphi_{i}\rangle$. We exploit $|\varphi_{i}\rangle=\int dp \varphi(p)|p\rangle$. The key thing here is to measure the physical quantity appearing in the interaction instead of its conjugate one. The measurement quantity communicates with the Hamiltonian of the system. If there is no postselection, it is a conserved quantity, and can not change at all. So its shift must be an artifact of postselection, which \emph{delivers}
 ``the correlations induced by the interaction between the system and the meter''. This choice for measurement is always feasible, at least in principle.

After interaction,  when $|\psi_{i,1}\rangle$ is selected (imaginary-value amplification), the whole system is $\int dp\frac{1}{\sqrt{2}}(e^{i\theta}e^{igp}|0\rangle+e^{-igp}|1\rangle)\varphi(p)|p\rangle$. A relative phase $2gp$ (the correlations) appears in the superposition state of the quantum system. Through the interaction, the meter system exerts a back-action to the quantum system. We should notice that, for pure state, entanglement still exists. When the final state of the quantum system $|\psi_{f,1}\rangle$ is selected, the final state of the meter system is
\begin{eqnarray}
\int dp\frac{1}{2P}(e^{i\theta}e^{igp}-e^{-igp})\varphi(p)|p\rangle=\int dp \varphi'(p) |p\rangle,
\end{eqnarray}
which must be equal to (1). Here interference still exists, but is not related to $\varphi(p)$. It seems that postselection changes the wavefunction of the meter system from $\varphi(p)$ to $\varphi'(p)$, so it has no classical meaning. 

Next we discuss the amplification effect. The mean momentum for $p$ in the final state is 
\begin{eqnarray}
\bar{p}_{f}&=&\int dp' \varphi'(p') \langle p' | \hat{p} \int dp \varphi'(p) |p\rangle \notag \\
&=&\int dp p |\varphi'(p)|^{2}  \notag \\
&=&\int dp p \frac{1}{2P}(1-\cos(\theta+2gp))|\varphi(p)|^{2}. 
\end{eqnarray} 
In this derivation, the interference disappears, which is different to reducing the mean position for $q$ in (1), that
\begin{eqnarray}
\bar{q}_{f}&=&\int dq' \varphi'(q') \langle q' | \hat{q} \int dq \varphi'(q) |q\rangle \notag \\
&=&\int dq q |\varphi'(q)|^{2}  \notag \\
&=&\int dq q \frac{1}{4P}(|\varphi(q+g)|^{2}-e^{i\theta}\varphi(q+g)\varphi'(q-g) \notag \\
&&-e^{-i\theta}\varphi'(q+g)\varphi(q-g)+|\varphi(q-g)|^{2}).
\end{eqnarray} 

The difference of the two results (4) and (5) are very interesting. For (5) the interference can not be removed (quantum or classical). However (4) can be derived from a purely classical one. Now the initial state of the system to be measured is prepared in a mixed state
 $\varrho_{i}=\int dp |\varphi(p)|^{2}|p\rangle \langle p|$. After interaction, we will have $\int dp |\psi(p)\rangle \langle \psi(p)|\otimes |\varphi(p)|^{2} |p\rangle \langle p| $, where $|\psi(p)\rangle=\frac{1}{\sqrt{2}}(e^{i\theta}e^{igp}|0\rangle+e^{-igp}|1\rangle)$. Here entanglement does not exist.  When the final state of the quantum system $|\psi_{f,1}\rangle$ is selected, the final state of the meter system is 
\begin{eqnarray}
\int dp \frac{1}{2P}(1-\cos(\theta+2gp))|\varphi(p)|^{2} |p\rangle \langle p| ,
\end{eqnarray} 
so the mean momentum $\bar{p}_{f}$ for the final state (6) is the same as that in (4). Considering the mean momentums $\bar{p}_{i}$ for the two initial states $|\psi_{i,1}\rangle$ and $\varrho_{i}$ are identical, the change of the mean momentum $\Delta p=\bar{p}_{f}-\bar{p}_{i}$ can not be distinguished between them.

When we choose $\varphi(q)$ or $\varphi(p)$, these means the measurement of $q$ and $p$ has been performed. Important is that, when measuring $p$, the entanglement between the quantum system and meter system is absolutely destroyed. Entanglement is not an indispensable component for imaginary-AWA (corresponding to imaginary-value amplification). When the final state $|\psi_{f,1}\rangle$ is selected, the probability distribution for measuring $p$ is $P_{1}(p,\theta)=|\frac{1}{2}(e^{i\theta}e^{igp}-e^{-igp})\varphi(p)|^{2}=\frac{1}{2}(1-\cos(\theta+2gp))|\varphi(p)|^{2}$. We can see the destructive interference of the probe system is only a part of the whole story, and the interference between preselection and postselection is the most important. Postselection induces a large change on the probability distribution of $p$ measurement.  When $\theta+2gp_{k}=2k\pi$, $k$ is an integer number, the corresponding success detection probability for $p_{k}$ is zero. So in this new perspective postselection plays a classical role for the meter change, although its realization here is quantum.

Ref. \cite{Howell15} did not mention the classicality. They  explicitly choose both of the final orthogonal states $|\psi_{f,2}\rangle$ and $|\psi_{f,3}\rangle$ for an almost balanced detection, which is an extension of Ref \cite{Bruder13}. A similar choice is also performed in \cite{Zeng15}. That is, they give up the traditional dark port postselection, and exploit new postselection. This is the reason why they claim ``without postselection'' (select ont from two). Most important thing in \cite{Howell15} is that they clarifies the meaning of weak amplification in this SB scheme via the difference of probability distribution of measuring $p$ at the two ports, which means that weak amplification and weak value are not one thing. This clarification is a significant progress for AWA. Thus the detection probability distribution for $p$ at two ports are respectively $P_{2}(p,\theta)=\frac{1}{2}(1-\sin(\theta+2gp))|\varphi(p)|^{2}$ and $P_{3}(p,\theta)=\frac{1}{2}(1+\sin(\theta+2gp))|\varphi(p)|^{2}$. The sum and difference of $P_{2}(p,\theta)$ and $P_{3}(p,\theta)$ are
\begin{eqnarray}
P_{+}(p,\theta)&=&P_{2}(p,\theta)+P_{3}(p,\theta)\notag \\
&=&|\varphi(p)|^{2},
\end{eqnarray}
and
\begin{eqnarray}
P_{-}(p,\theta)&=&-(P_{2}(p,\theta)-P_{3}(p,\theta))  \notag \\
&=&|\varphi(p)|^{2}\sin(\theta+2gp).
\end{eqnarray}
$P_{+}(p,\theta)$ corresponds to the probability distribution of $p$ in the initial state. The amplification effect results from the difference signal (2), which can be  remarkably different from the sum one (1).

Thus the AWA in the SB scheme has a classical root, which is different from the (real-value) amplification in AAV scheme. When $\theta+2gp_{k}=2k\pi$, $k$ is an integer number, the corresponding success detection probability for $p_{k}$ is zero. This situation is the same as that in AAV scheme. The key component in AWA is to largely change the output signal. Although the interaction between the quantum system and the meter system is very weak, we can also observe it via postselection, which can be used in observing tiny effect.

The results in \cite{Wang15,Howell15} are beyond our expectation. Now we have two AWA schemes. From the view in \cite{Howell15}, we can easily improve the traditional AAV scheme, which is also done in \cite{Zeng15}. We also detect the bright port $|\psi_{f,4}\rangle=\frac{1}{\sqrt{2}}(|0\rangle+|1\rangle)$ and its detection probability for $p$ is $P_{4}(p,\theta)=\frac{1}{2}(1+\cos(\theta+2gp))|\varphi(p)|^{2}$. So $P_{+}(p,\theta)=P_{1}(p,\theta)+P_{4}(p,\theta)=|\varphi(p)|^{2}$ and an amplification signal can be defined as
\begin{eqnarray}
P'_{-}(p,\theta)&=&P_{+}(p,\theta)-(P_{1}(p,\theta)-P_{4}(p,\theta))  \notag \\
&=&|\varphi(p)|^{2}(1-\cos(\theta+2gp)).
\end{eqnarray}
In this improved scheme all of the information are used.

We can generalize the idea to arbitrarily orthogonal postselections, $|\psi(\chi)_{f,+}\rangle=\frac{1}{\sqrt{2}}(|0\rangle+e^{i\chi}|1\rangle)$ and $|\psi(\chi)_{f,-}\rangle=\frac{1}{\sqrt{2}}(|0\rangle-e^{i\chi}|1\rangle)$, where $\chi\in [0, \frac{\pi}{2}]$ (This arbitrary postselection has been also done for joint weak measurements in \cite{Zeng15}).
So the detection probability for $p$ are
\begin{eqnarray}
Pr_{\pm}(p,\theta)&=&\frac{1}{2}(1\pm \cos(\theta+2gp+\chi))|\varphi(p)|^{2}.
\end{eqnarray}
When $\chi=0$, it corresponds to the AAV scheme. When $\chi=\frac{\pi}{2}$, it corresponds to the SB scheme. The amplification signal can be defined as
\begin{eqnarray}
Pr(p,\theta)&=&\cos(\chi)P_{+}(p,\theta)-(Pr_{+}(p,\theta)-Pr_{-}(p,\theta))  \notag \\
&=&((1- \cos(\theta+2gp))\cos(\chi)  \notag \\
&&+ \sin(\theta+2gp)\sin(\chi))|\varphi(p)|^{2}
\end{eqnarray}
This is the general result for AWA.

\begin{figure}[b]
\includegraphics[scale=0.14]{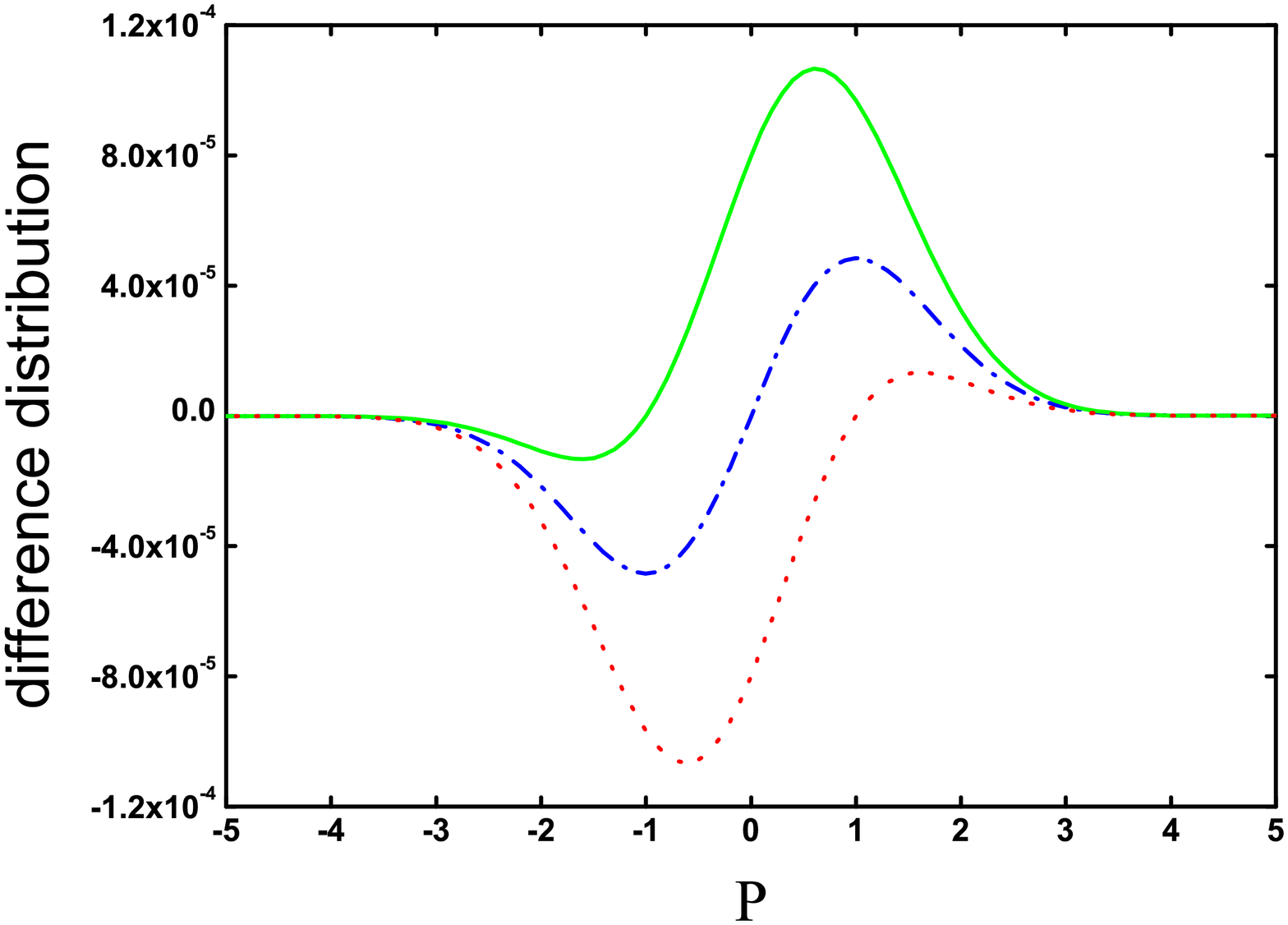}.
\includegraphics[scale=0.14]{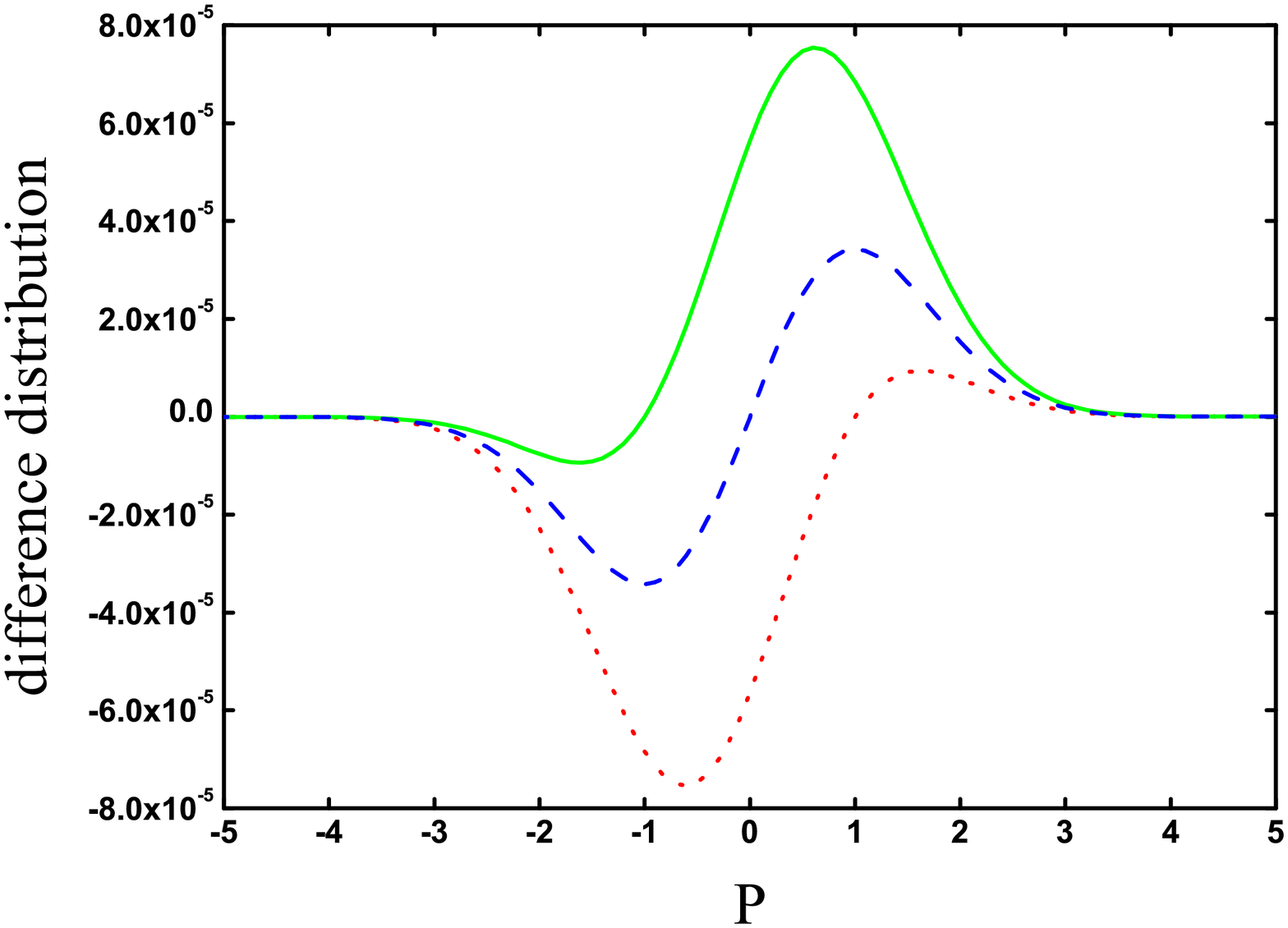}.
\includegraphics[scale=0.14]{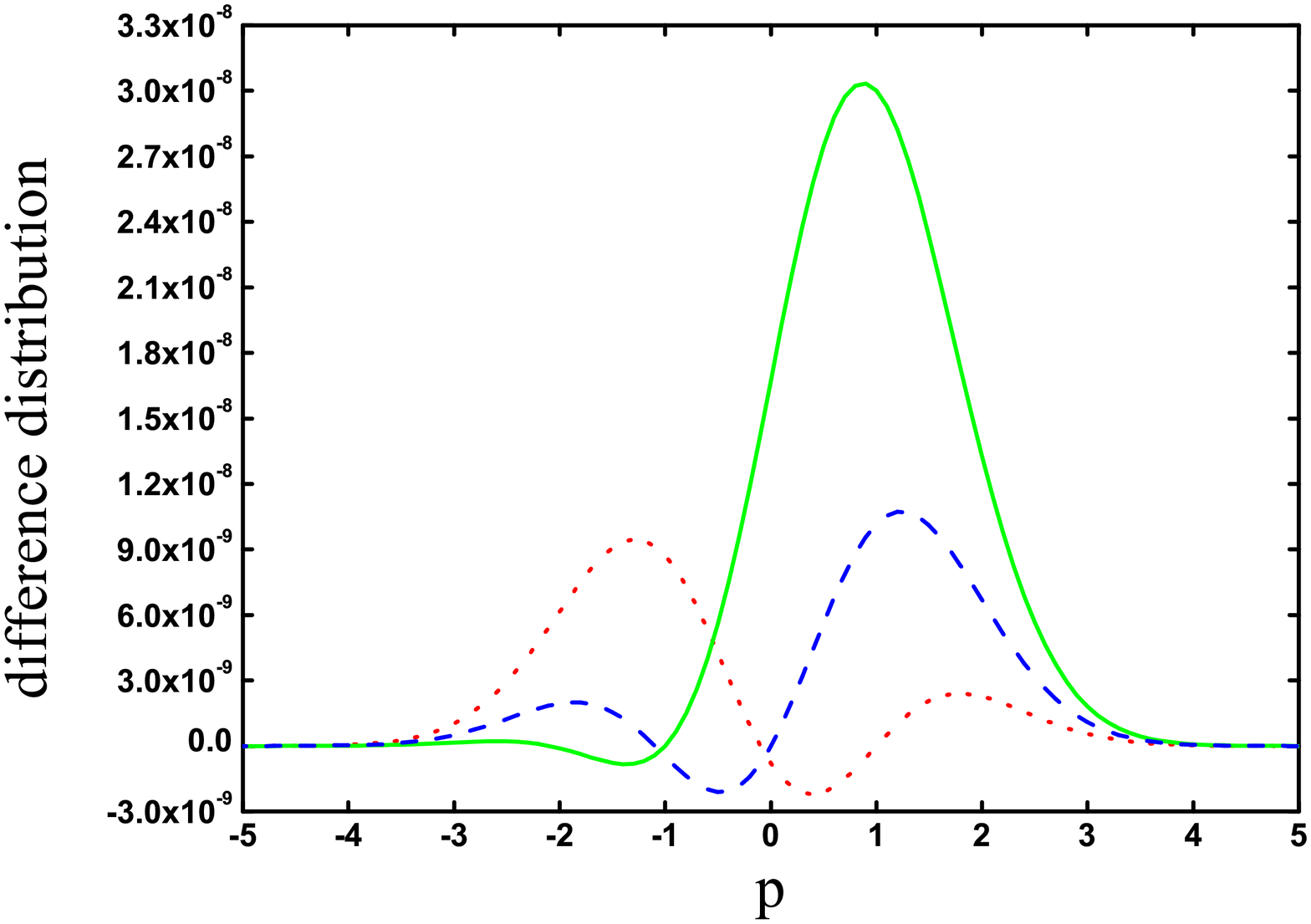}.
\includegraphics[scale=0.14]{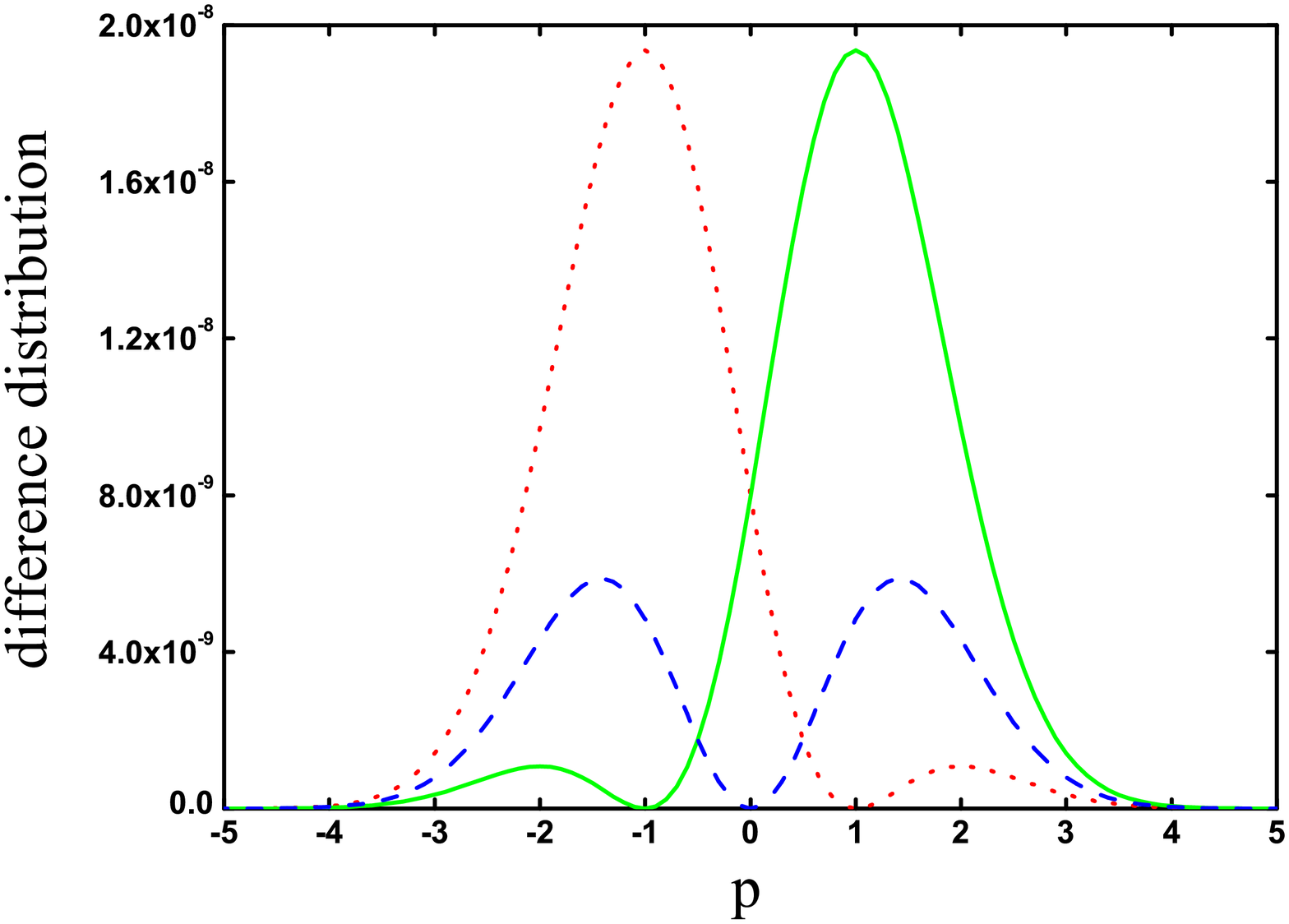}.
\caption{(a), (b), (c), (d) plots the difference distribution $Pr_{f}(p,\theta)$ for $\chi=\frac{\pi}{2}$, $\chi=0.5\times \frac{\pi}{2}$, $\chi=0.00007\times \frac{\pi}{2}$, $\chi=0$, where $\theta=0.0002$ (green-solid line),  $\theta=0$ (blue-dashed line) and $\theta=-0.0002$ (red-dotted line). }
\end{figure}

Next we generalize the result to any meter state. Ref. \cite{coherent} provides a coherent-state view for AWA, which bases on the fact that any density matrix could be expanded diagonally in terms of coherent states \cite{Glauber,Sudarshan}. Any meter system can be described as $\rho_{i}=\int d^{2}\alpha P(\alpha)|\alpha\rangle \langle \alpha |$. The weight function $P(\alpha)$ is known as the Glauber-
Sudarshan $P$-distribution. The whole system is prepared as $|\psi_{i,1}\rangle\langle \psi_{i,1} |\otimes \rho_{i}$. After interaction we have
\begin{eqnarray}
\rho_{qm}&=&\frac{1}{2}\int d^{2}\alpha P(\alpha)(e^{i\theta}|0\rangle e^{ig\hat{p}}|\alpha\rangle+|1\rangle e^{-ig\hat{p}}|\alpha\rangle)  \notag \\
&&(e^{-i\theta}\langle 0| \langle\alpha|e^{-ig\hat{p}}+\langle 1| \langle\alpha|e^{ig\hat{p}})
\end{eqnarray}
When the final states of the quantum system $|\psi(\chi)_{f,\pm}\rangle$ are selected, the final state of the meter is
\begin{eqnarray}
\rho_{f,\pm}&=&\frac{1}{4}\int d^{2}\alpha P(\alpha)(e^{i\theta} e^{ig\hat{p}}|\alpha\rangle\pm e^{-i\chi}e^{-ig\hat{p}}|\alpha\rangle)  \notag \\
&&(e^{-i\theta} \langle\alpha|e^{-ig\hat{p}}\pm e^{i\chi}\langle\alpha|e^{ig\hat{p}})
\end{eqnarray}
When we measure $p$, the trick occurs. The probability distribution for $p$ is
\begin{eqnarray}
\langle p |\rho_{f,\pm}| p \rangle&=&\frac{1}{4}\int d^{2}\alpha P(\alpha)(e^{i\theta} e^{igp}\langle p|\alpha\rangle\pm e^{-i\chi}e^{-igp}\langle p|\alpha\rangle)  \notag \\
&&(e^{-i\theta} \langle\alpha|p \rangle e^{-igp}\pm e^{i\chi}\langle\alpha|p \rangle e^{igp}) \notag \\
&=&\frac{1}{2}(1\pm \cos(\theta+2gp+\chi))\langle p |\rho_{i}| p \rangle
\end{eqnarray}
Thus we have amplification signal
\begin{eqnarray}
Pr_{f}(p,\theta)&=&((1- \cos(\theta+2gp))\cos(\chi)  \notag \\
&&+ \sin(\theta+2gp)\sin(\chi))P_{i}(p).
\end{eqnarray}
Here $P_{i}(p)=\langle p |\rho_{i}| p \rangle$ is the probability distribution for $p$ in the initial state of the probe system, so the general result holds for any meter system. When $\theta+2gp_{k}=2k\pi$, $k$ is an integer number, the corresponding detection probability for measuring $p_{k}$ is zero. In a word, postselection changes the probability distribution of $p$ significantly.

We use a Gaussian state distribution $P_{i}(p,\theta)=\frac{1}{\sqrt{2\pi}\sigma}e^{-\frac{p^{2}}{2\sigma^{2}}}$, where $\sigma$ is the width of the initial state. We adopt $g=0.0001$. Fig. 1 (a), (b), (c), (d) plots the difference distribution $Pr_{f}(p,\theta)$ for $\chi=\frac{\pi}{2}$, $\chi=0.5\times \frac{\pi}{2}$, $\chi=0.00007\times \frac{\pi}{2}$, $\chi=0$, where $\theta=0.0002$ (green-solid line),  $\theta=0$ (blue-dashed line) and $\theta=-0.0002$ (red-dotted line). We can see, when amplification effect becomes larger (from 0.5$\sigma$ to $\sigma$), the difference distribution becomse much smaller (from $1.1\times 10^{-4}$ to $2.0\times 10^{-8})$.

In conclusions classicality and amplification in postselected weak measurement is clarified and generalized. For imaginary weak amplification, postselection plays a classical role, and this amplification can be obtained from arbitrary orthogonal postselections. These results can enhance our understanding of weak measurement and its application in sensing, precision and gravitational wave detection.

\end{document}